\documentclass[preprint,onecolumn,preprintnumbers,amsmath,amssymb]{revtex4}
\usepackage{txfonts}
\usepackage{dsfont}
\usepackage{amssymb}
\usepackage{booktabs}
\usepackage{graphicx}
\usepackage{color}
\usepackage{dcolumn}
\usepackage{amsmath}
\usepackage{epstopdf}
\usepackage{bm}
\usepackage{ulem}
\hfuzz=\maxdimen
\begin{document}
\title{Three consecutive quantum anomalous Hall gaps in a metal-organic network}
\author{Xiang-Long Yu$^{1,2}$}
\author{Tengfei Cao$^{3}$}
\author{Rui Wang$^{4}$}
\author{Ya-Min Quan$^{5}$}
\author{Jiansheng Wu$^{1,2,6}$}
\email[Corresponding author. E-mail: ]{wujs@sustech.edu.cn}
\affiliation{1 Shenzhen Institute for Quantum Science and Engineering, Southern University of Science and Technology, Shenzhen 518055, China}
\affiliation{2 International Quantum Academy, Shenzhen 518048, China}
\affiliation{3 State Key Laboratory of Solidification Processing, Center for Advanced Lubrication and Seal Materials, School of Materials Science and Engineering, Northwestern Polytechnical University, Xi'an, Shanxi 710072, China}
\affiliation{4 Institute for Structure and Function, Department of Physics and Center for Quantum Materials and Devices, Chongqing University, Chongqing 400044, China}
\affiliation{5 Key Laboratory of Materials Physics, Institute of Solid State Physics, HFIPS, Chinese Academy of Sciences, Hefei 230031, China}
\affiliation{6 Guangdong Provincial Key Laboratory of Quantum Science and Engineering, Southern University of Science and Technology, Shenzhen 518055, China}


\begin{abstract}
 In the quantum anomalous Hall (QAH) effect, chiral edge states are present in the absence of magnetic fields due to the intrinsic band topology. In this work, we predict that a synthesized two-dimensional metal-organic material, a Fe(biphenolate)$_3$ network, can be a unique QAH insulator, in which there are three consecutive nontrivial bandgaps. Based on first-principles calculations with effective model analysis, we reveal such nontrivial topology is from the $3$d$_{xz}$ and $3$d$_{yz}$ orbitals of Fe atoms. Moreover, we further study the effect of substrates, and the results shows that the metallic substrates used in the experiments (Ag and Cu) are unfavorable for observing the QAH effect whereas a hexagonal boron nitride substrate with a large bandgap may be a good candidate, where the three consecutive QAH gaps appear inside the substrate gap. The presence of three consecutive bandgaps near the Fermi level will significantly facilitate observations of the QAH effect in experiments.
\end{abstract}

\maketitle

\section{Introduction}
Topological systems have attracted much attention due to their unique physics and novel properties, \textit{e.g.}, nontrivial bulk topologies, edge/boundary states, Weyl fermions, and Majorana zero modes. Besides their fundamental physical importance, they also provide promising platforms for applications in quantum information and computing as well as spintronic devices.
For example, Majorana zero modes that obey non-Abelian statistics can be used as robust building blocks in quantum computing, and topologically protected edge states can be used to realize ideal transport without dissipation because of the absence of back-scattering~\cite{RMP87.137,RMP80.1083}.

Recently, metal-organic topological materials have been extensively studied due to their unique advantages over inorganic counterparts, \textit{e.g.}, long spin coherence time and high tunability in both structural symmetry and charge doping~\cite{CM16.4413,JPDAP40.R205,ACR54.416}.
To date, several metal-organic topological materials have been proposed and present novel quantum phenomena, such as the following:
\begin{itemize}
	\item Quantum spin Hall effect in
Bi$_2$(C$_6$H$_4$)$_3$~\cite{NC4.1471},
Pb$_2$(C$_6$H$_4$)$_3$~\cite{NC4.1471,PRB93.041404},
Ni$_3$(C$_6$S$_6$)$_2$~\cite{NL13.2842,JACS136.14357},
and Al$_2$(C$_6$O$_4$Cl$_2$)$_3$~\cite{PCCP22.25827}
  \item Quantum anomalous Hall (QAH) effect in In$_2$(C$_6$H$_4$)$_3$~\cite{PRL110.106804},
Tl$_2$(C$_6$H$_4$)$_3$~\cite{APL112.033301},
Mn$_2$(C$_6$H$_4$)$_3$~\cite{PRL110.196801,NRL9.690},
and Cu$_2$(C$_8$N$_2$H$_4$)$_3$~\cite{NR13.1571}
  \item flat-band characteristic in Cu$_3$(C$_6$N$_6$H$_6$)$_2$~\cite{Nanoscale11.955},
Au-phenalenyl organometallic frameworks~\cite{PRB94.081102R},
and sp$^2$ carbon-conjugated covalent organic frameworks, which can be viewed as an assembly of pyrene and a $1$,$4$-bis(cyanostyryl)benzene ligand~\cite{NC10.2207,NC11.66,Science357.673},
and phthalocyanine-based metal-organic frameworks~\cite{NL20.1959}
\end{itemize}
However, few of these materials have been successfully synthesized and fewer have exhibited topologically nontrivial properties experimentally. Hence, investigating nontrivial topologies with synthesized materials instead of making theoretical predictions about new topological materials may be more efficient.

In this work, we study a synthesized two-dimensional ($2$D) metal-organic material (MOM), the Fe(biphenolate)$_3$ network with the chemical formula Fe$_2$O$_6$(C$_6$H$_4$)$_6$, which was first synthesized on the substrates Ag($111$) and Cu($100$) by Stepanow \textit{et al.}~\cite{ACIE46.710}. We calculate its electronic and topological properties and find a topologically nontrivial ferromagnetic (FM) state exhibiting the QAH effect.
Interestingly, three consecutive QAH gaps were obtained, which have not been found in other organic materials and will be very useful and advantageous for experimental observations.
When the electronic correlation is considered, the QAH states are robust within the bulk bandgaps.
We also study and discuss the effect of substrates on the topological properties of this $2$D MOM, including the metal substrates that were used to synthesize the experimental samples and three insulating ones, Al$_2$O$_3$, SiO$_2$, and hexagonal BN (\textit{h}-BN). Our calculation results show that the \textit{h}-BN substrate may be a good candidate for producing the QAH effect and that the nontrivial QAH states occur within the bandgap of the substrate as expected. These results may provide useful information for experimental measurements of the topological properties of $2$D MOMs.

\section{Quantum anomalous Hall effect}
\begin{figure*}
\centering
\includegraphics[width=15.0 cm]{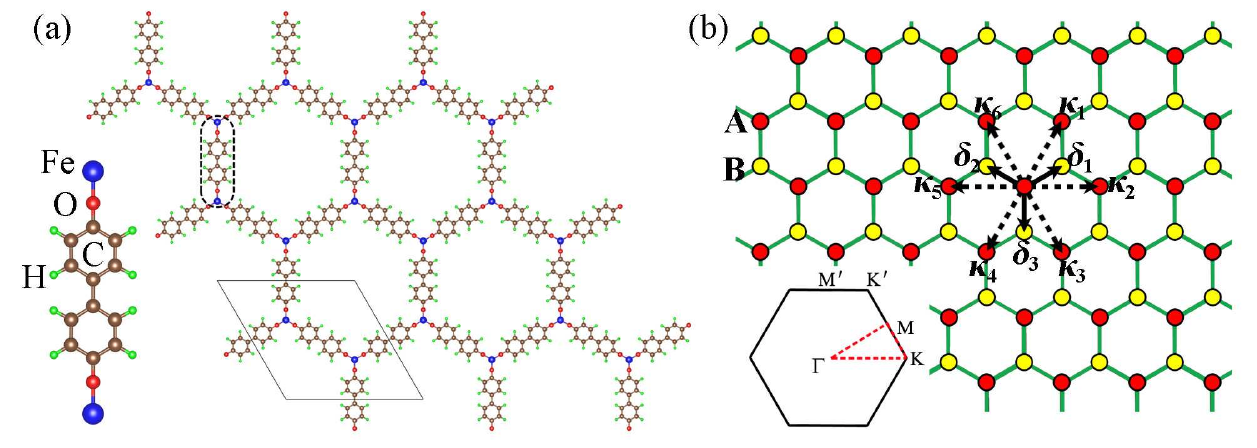}
\caption{$2$D lattice structure. (a) Schematic atomic structure of the Fe(biphenolate)$_3$ network. Blue, red, brown, and green spheres denote Fe, O, C, and H atoms, respectively. The solid rhombus denotes the unit cell. The inset is an enlargement of the molecules in the dashed box. (b) Honeycomb lattice formed by Fe atoms, which is divided into A and B sublattices. The inset shows the corresponding first Brillouin zone with high-symmetry points and paths.}
\label{fig:struct}
\end{figure*}

Figure~\ref{fig:struct}(a) shows the $2$D atomic structure of the Fe(biphenolate)$_3$ network. There are 68 atoms in each unit cell: two Fe, six O, 36 C, and 24 H atoms. The Fe atoms form a honeycomb lattice. Two nearest-neighbor Fe atoms are connected by a $4$,$4'$-biphenol ligand. If we regard the ligand as a lattice site, there is a Kagome lattice on the $2$D plane. Overall, there is a honeycomb-Kagome lattice~\cite{JPCC122.18659}.

\begin{figure*}
  \centering
  \includegraphics[width=1.0 \columnwidth]{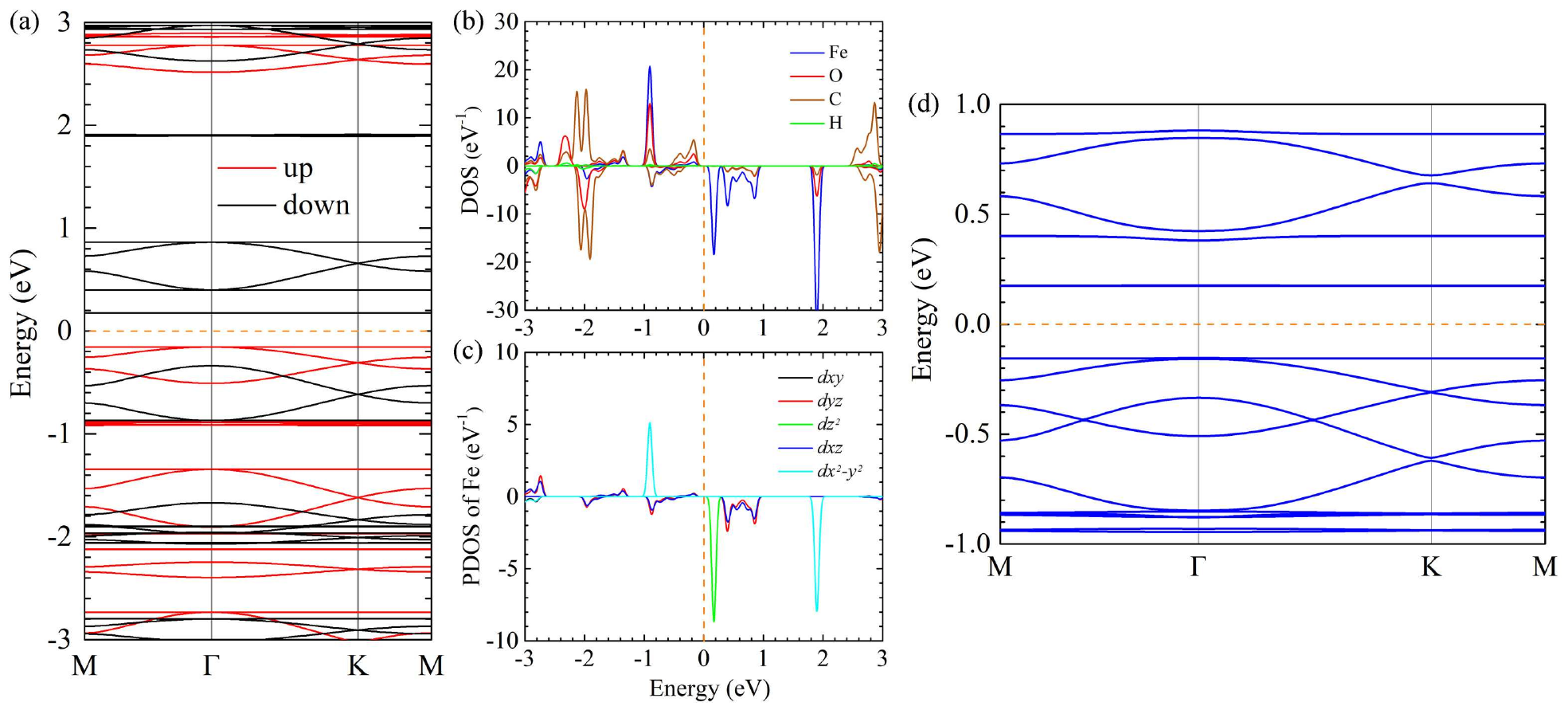}
  \caption{Electronic structures of the Fe(biphenolate)$_3$ network. (a) Band structure, (b) PDOS of the four elements, and (c) PDOS of the five Fe-$3$d orbitals of the FM state without SOC. Positive and negative values of PDOS correspond to up and down spins, respectively. (d) Band structure of the FM state with SOC.}
  \label{fig:Band_DOS_FM}
\end{figure*}

Our calculations show that Fe(biphenolate)$_3$ has various magnetic configurations with rich properties, especially the nontrivial topology in the FM state. Here, we focus on the FM state, and other magnetic configurations with trivial topology and the comparison among them are provided in the Supplementary Information~\cite{SM}.
We calculate the electronic structures of the FM state after optimizing the crystal.
The magnetic moment of each Fe atom is $3.62\mu_B$.
Figure~\ref{fig:Band_DOS_FM}(a) shows the band structure without considering SOC or electronic correlation. The system is a FM insulator. There is an energy gap of about $2.7$~eV for the up spin, and an energy gap of about $0.5$~eV for the down spin. Some bands are very flat, suggesting strong localization.
Figures~\ref{fig:Band_DOS_FM}(b) and~\ref{fig:Band_DOS_FM}(c) show that the projected density of states (PDOS) near the Fermi level is mainly from Fe, O, and C atoms, especially Fe-3d orbitals.
Notably, there is an isolated flat band above the Fermi energy level, which is due to  the Fe-3d$_{3z^2-r^2}$ orbital. The reason for the flat band is that the 3d$_{3z^2-r^2}$ orbital is not bonded to other orbits in this flat $2$D system, leading to a perfect energy level.
Besides, there are four bands due to Fe-3d$_{yz}$ and 3d$_{xz}$ orbitals in the range $0.4$--$0.9$~eV.
Since the material has a $C_3$ symmetry, the two orbitals are completely degenerate and so are Fe-3d$_{xy}$ and 3d$_{x^2-y^2}$ orbitals.

Figure~\ref{fig:Band_DOS_FM}(d) shows the FM band structure with SOC. The energy gap is about $0.3$~eV. Note that the SOC interaction obviously opens three energy gaps between the four bands due to the degenerate orbitals Fe-3d$_{yz}$ and 3d$_{xz}$, which are $22.6$, $35.6$, and $17.8$~meV, respectively. This possibly leads to a nontrivial topology. To verify this, we use \textsc{Wannier}$90$ to fit the bands according to the DFT results~\cite{CPC178.685} (see the Supplementary Information for more details~\cite{SM}). We also calculate the surface states of a semi-infinite system using an iterative Green function in the software \textsc{WannierTools}~\cite{CPC224.4059}.

\begin{figure*}
  \centering
  \includegraphics[width=1 \columnwidth]{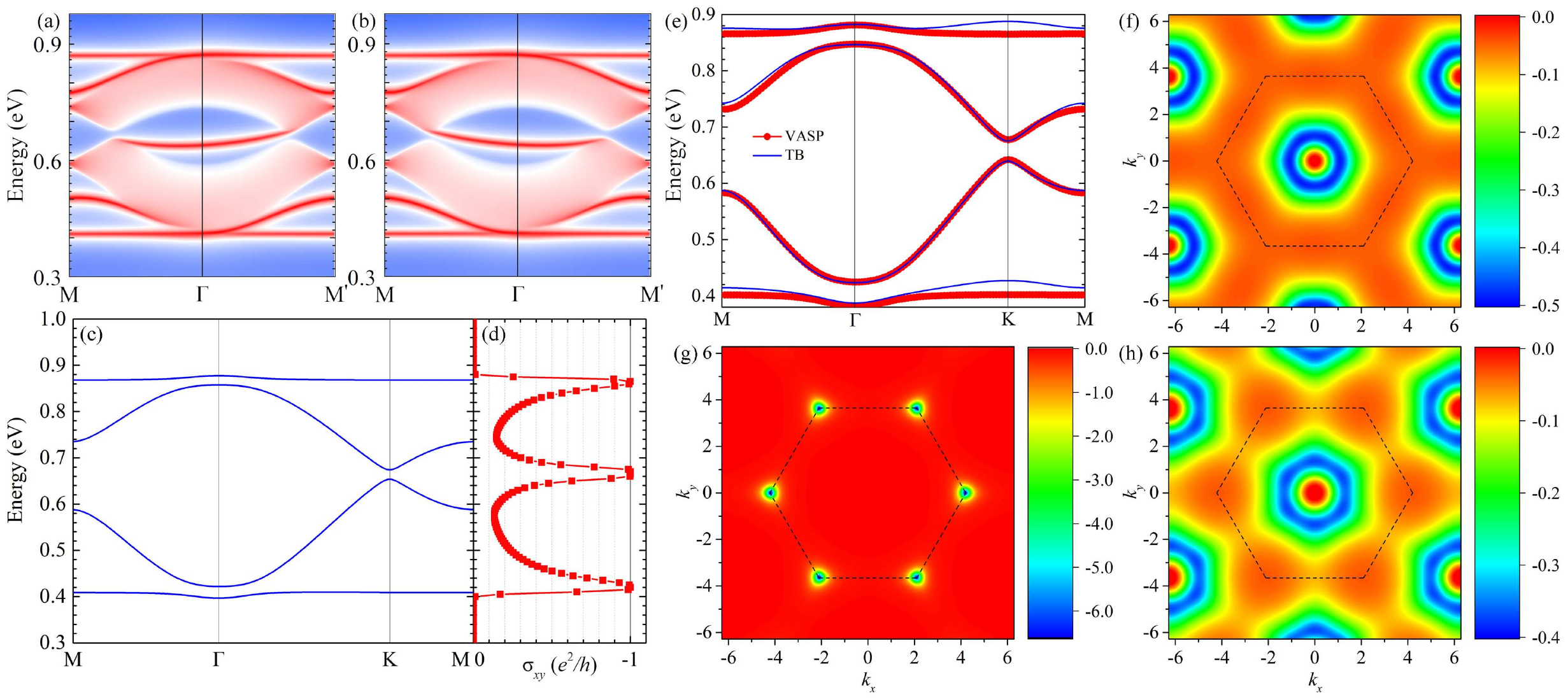}
  \caption{Topological properties of the Fe(biphenolate)$_3$ network. Band structures of semi-infinite systems with (a) left and (b) right edges. The high (low) density of states is colored red (blue). (c) Four bulk bands of the FM state with SOC. (d) Hall conductivity of the FM state.
  (e) Band structures from DFT and tight-binding calculations. Berry curvature in the Brillouin zone when the Fermi level is within the (f) first, (g) second, or (h) third gaps from bottom to top. The black dashed box is the first Brillouin zone.}
  \label{fig:topo_prop}
\end{figure*}

Figures~\ref{fig:topo_prop}(a) and~\ref{fig:topo_prop}(b) show the band structures of semi-infinite systems with left and right edges, respectively. By comparing them with the bulk band structure, we can see that there are edge states within the three energy gaps, corresponding to a nontrivial topology. For the left edge state, the slope of the band is positive, suggesting that the state propagates in the positive direction. In contrast, the right edge state propagates in the negative direction. This is due to the destroyed time-reversal symmetry. Therefore, the FM state is a QAH insulator when the Fermi level shifts into any one of the three gaps.
For further confirmation, we calculate the Hall conductivity, which has perfect quantum platforms within the three bulk gaps, as plotted in Figs.~\ref{fig:topo_prop}(c) and~\ref{fig:topo_prop}(d). The value is $-e^2/h$, corresponding to QAH edge states.

We construct an effective four-band tight-binding model in the Methods part to fit the four DFT bands with the three consecutive QAH gaps. There is no definitive strategy for the fitting. We tried to achieve consistency between the DFT and the model bands within a reasonable parameter range. The fitting results are plotted in Fig.~\ref{fig:topo_prop}(e) with fitting parameters $t = 0.153$~eV, $t' = -0.005$~eV, $\lambda = 0.018$~eV, and ${\varepsilon _0} = 0.650$~eV. The results from both methods agree well with each other, especially for the middle two bands. Note that the next-nearest-neighbor (NNN) hopping parameter $t'$ is important despite that $|t'| \ll |t|$. This is because the DFT results show that the four bands are not symmetric with respect to the middle gap. In the tight-binding model, breaking the symmetry requires the NNN hopping.

Although the DFT calculations indicate that the three gaps are topologically nontrivial, they cannot determine whether the nontrivial topology is completely attributed to the two orbitals Fe$-3d_{xz}$ and 3d$_{yz}$ because the four bands are also partly due to the O-$2$p orbitals, as shown in Fig.~\ref{fig:Band_DOS_FM}(b). All the bands under the considered gap are included in principle when calculating the topological properties, such as edge states and Hall conductivity. Therefore, the effective tight-binding model is helpful for clarifying this issue. Based on the fitting result, we calculate the Berry curvatures and plot them in Figs.~\ref{fig:topo_prop}(f)--\ref{fig:topo_prop}(h). When the Fermi level is within the first and third gaps, the main contribution is from the point $\Gamma$ and its surroundings. When the Fermi level is within the second gap, the Berry curvature is strictly localized to the points $K$ and $K'$. We can obtain the Chern number $C$ by integrating the Berry curvature over the Brillouin zone. All three cases give $C = -1$, exactly consistent with the DFT calculation result. We can also confirm that the nontrivial topology is completely due to the Fe$-3d_{xz}$ and 3d$_{yz}$ orbitals.

\begin{figure*}
  \centering
  \includegraphics[width=1 \columnwidth]{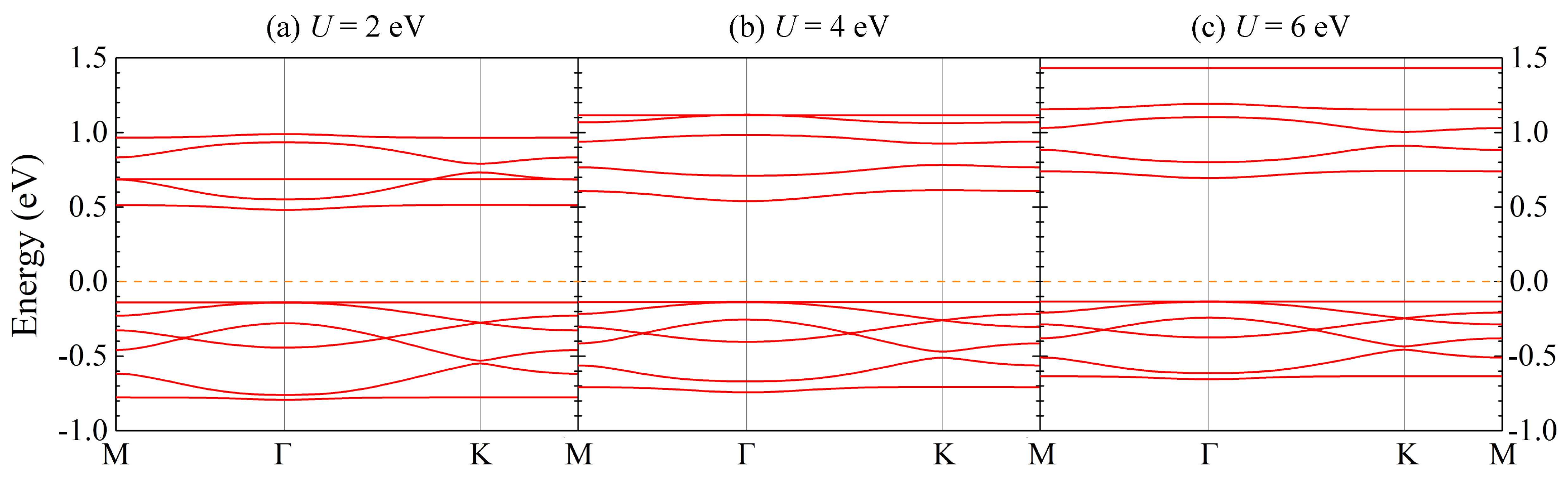}
  \caption{Band structures of the FM state with SOC and $U =$ (a) $2$~eV, (b) $4$~eV, or (c) $6$~eV.}
  \label{fig:band246eV}
\end{figure*}

The electronic correlation has to be taken into account for an insulating material containing a $3$d element. For the nontrivial FM state, we consider different local Coulomb interactions from $U = 0$ to $6$~eV and investigate its electronic and topological properties.
Figure~\ref{fig:band246eV} shows the band structures with SOC and $U = 2$, $4$, or $6$~eV.
Notice that the flat band from the Fe-3d$_{3z^2-r^2}$ orbital increases with the electronic correlation. But since this orbital does not form any bonds with other orbitals, as we mentioned above, we can regard it as a decoupled level and neglect it during the current analysis.
In addition, with increasing $U$, the bands due to Fe-3d orbitals become narrower and flatter, especially the four bands near the Fermi level. The bandgaps also become wider. Moreover, the symmetry of the $2$D material is not broken. Thus, the topology does not change when the electronic correlation is considered. This suggests that the system with nonzero electronic correlation is still topologically nontrivial.

\section{Effect of substrates}
Most theoretical predictions of metal-organic topological systems have been based on calculations for freestanding $2$D materials. However, the substrate usually has a key role in the synthesis and applications of a $2$D material. The choice of substrate can change the intrinsic structural symmetry and the physical properties.
Here, we study several substrates and discuss their effect on the properties of the Fe(biphenolate)$_3$ network. The substrates are Ag($111$), Cu($100$), Al$_2$O$_3$($0001$), SiO$_2$($111$), and \textit{h}-BN. Their space group, bulk gap, lattice parameter, and mismatch are summarized in Table~\ref{tab:substrate}.

\begin{table*}
\caption{Five substrates for the Fe(biphenolate)$_3$ network~\cite{materialsproject}. $D_{NN}$ is defined as the distance between nearest-neighbor (NN) atoms on the substrate surface.}
\begin{tabular}{cccccc}
\hline
\hline
Substrate & Bulk Space Group & Bulk Gap & Lattice Parameter of Substrate & $D_{NN}$ & Mismatch \\
&& (eV) & (\AA) & (\AA) & (\%) \\
\hline
Ag($111$)               & Fm\={3}m & 0   & 2.89 & 2.89 & 1.0 \\
Cu($100$)               & Fm\={3}m & 0   & 3.62 & 2.56 & -- \\
Al$_{2}$O$_{3}$($0001$) & R\={3}c  & 6.5 & 4.76 & 4.76 & 1.9 \\
SiO$_2$($111$)          & FD\={3}m & 6.0 & 5.06 & 5.06 & 7.7 \\
\ \ \textbf{\textit{h}-BN}          & \textbf{P6{3}/mmc} & \textbf{4.3} & \textbf{2.51} & \textbf{1.45} & \textbf{3.3} \\
\hline
\hline
\end{tabular}
\label{tab:substrate}
\end{table*}

We first investigate the effect of substrates Ag($111$) and Cu($100$), which have been used in experiments on the Fe(biphenolate)$_3$ network~\cite{ACIE46.710}.
The Ag($111$) substrate has a lattice parameter of $a=2.89$~\AA{}, leading to a small mismatch of $1\%$ for the Fe(biphenolate)$_3$ network. The Cu($100$) surface has fourfold symmetry. The hexagonal network cannot perfectly adsorb at the substrate, leading to some deformed networks, as shown in experiments~\cite{ACIE46.710}.
However, note that both Ag and Cu are good conductors, and there are no bandgaps near the Fermi level.
Our calculation results for the Ag($111$) substrate provided in the Supplementary Information show that although the chiral structure observed in experiments is obtained here, the electronic states of the Fe(biphenolate)$_3$ network are completely covered by those of the substrates near the Fermi level, so that the nontrivial topology may not be accurately measured~\cite{SM}. Thus, the Ag($111$) and Cu($100$) substrates are unfavorable for producing the QAH effect.

Naturally, a substrate with a large bandgap is required to produce the nontrivial topology.
Besides, from the results of the Ag($111$) substrate, we learned that large interstices between the surface atoms of the substrates should be avoided, as these can cause an uneven Fe(biphenolate)$_3$ network. For example,
Al$_2$O$_3$ and SiO$_2$ are insulators with large bandgaps (${>} 6$~eV). The Al$_2$O$_3$(0001) and SiO$_2$(111) surfaces have threefold symmetry with a matching lattice, as presented in Table~\ref{tab:substrate}.
However, we do not think they can be good candidates because of the large interstices between their surface atoms.
When Fe atoms fall into the interstices, the nontrivial QAH states due to the Fe-3d$_{xz}$ and 3d$_{yz}$ orbitals are completely destroyed.

\begin{figure*}
  \centering
  \includegraphics[width=0.9 \columnwidth]{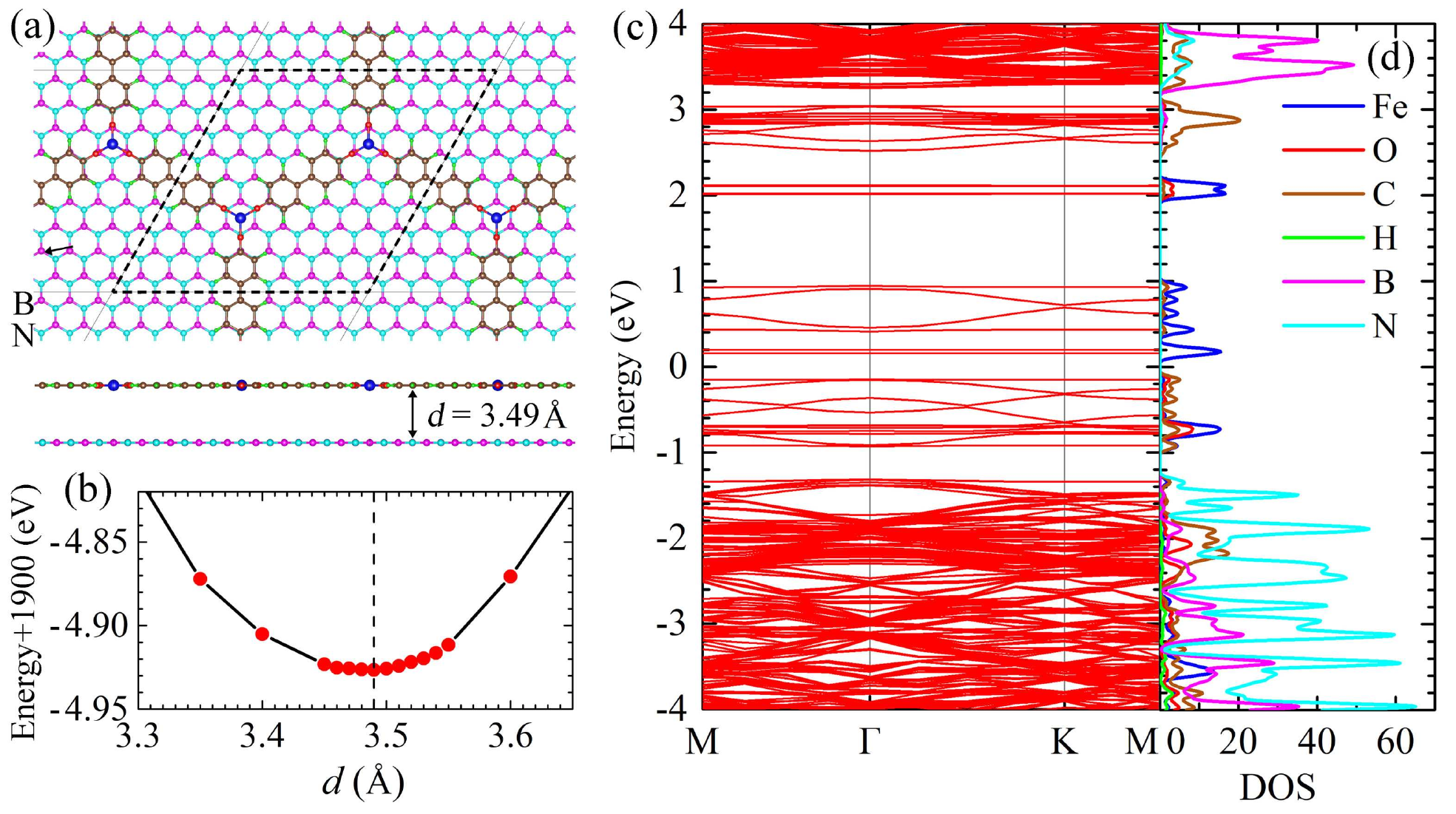}
  \caption{Effect of the \textit{h}-BN substrate on lattice and electronic structures. (a) Top and side views of the atomic configuration of the Fe(biphenolate)$_{3}$/\textit{h}-BN hybrid system. The dashed box is the unit cell. (b) Total energy as a function of the interlayer distance $d$. (c) Band structure and (d) PDOS of the hybrid system.}
  \label{fig:BN_substrate}
\end{figure*}

To avoid these situations, we consider boron nitride compounds with a smaller atomic distance.
They have various crystalline polymorphs analogous to carbon, such as \textit{h}-BN, diamond-like cubic BN, and onion-like fullerenes~\cite{CSR43.934}. Among them, \textit{h}-BN is the most stable allotrope thermodynamically, with strong sp$^2$ covalent bonds in the plane and weak van der Waals interactions between layers. Thus, it can act as a stable substrate or buffer layer~\cite{PSS96.100615,PE56.43,CSR43.6537,SR6.20152}.
Importantly, it provides a matching lattice for the Fe(biphenolate)$_3$ network, and the mismatch of $3.3\%$ is in an acceptable range. In addition, a theoretical calculation gives a bulk gap of ${\sim} 4.3$~eV~\cite{materialsproject}. Actually, the bandgap obtained from first-principles calculations is underestimated with respect to the experimental value ($6.07$~eV)~\cite{NL12.161}.
After geometry optimization, the Fe(biphenolate)$_3$ network remains almost flat and the distance to the \textit{h}-BN substrate is $d = 3.49$~\AA{}, as shown in Figs.~\ref{fig:BN_substrate}(a) and~\ref{fig:BN_substrate}(b).
The band structure plotted in Fig.~\ref{fig:BN_substrate}(c) is like that of the freestanding system near the Fermi level. The density of states in Fig.~\ref{fig:BN_substrate}(d) also shows that using \textit{h}-BN leads to a bandgap of ${\sim} 4$~eV. In particular, the four bands due to the Fe-3d$_{xz}$ and 3d$_{yz}$ orbitals in the range $0.2$--$1.0$~eV are not affected by the substrate.
There are three consecutive gaps ($43.6$, $23.9$, and $34.6$~meV), and these are close to those of the freestanding case, indicating that the QAH states still exist within the three gaps. Therefore, the \textit{h}-BN substrate may be a good candidate for producing the QAH effect.

\section{Discussion}
In the related experiments, $2$D chirality was observed such that the biphenolate ligands did not point toward the central Fe atom but rather tilted clockwise or anticlockwise~\cite{ACIE46.710}. However, our structural optimization of the Fe(biphenolate)$_3$ monolayer without any substrates did not produce chirality. Hence, we artificially rebuilt the lattice structure. The Fe-O-C angle was changed to $150.24^\circ$, and all the bond lengths were kept the same as those in the optimized structure. Repeating the calculations, we obtained qualitatively similar results, including the electronic and topological properties (see the Supplementary Information for more details~\cite{SM}). In particular, there were still three consecutive QAH gaps.

Notice that the $2$D transition-metal oxide V$_2$O$_3$ has a similar lattice (the honeycomb-Kagome lattice) and band structures (four nontrivial bands), and has also been theoretically predicted to be a QAH insulator~\cite{PRB95.125430,JMMM564.170161}. However, first-principles results show that both the electronic correlation $U$ and SOC must be considered simultaneously. A nontrivial gap does not occur when only SOC is taken into account without $U$. Besides, although V$_2$O$_3$ and the Fe(biphenolate)$_{3}$ network have similar energy bands for the QAH effect due to the 3d$_{xz}$ and 3d$_{yz}$ orbitals, the former does not have three QAH gaps. Its two gaps are covered by the V-3d$_{xy, x^2-y^2}$ and O-p electronic states and only the middle gap survives. Therefore, based on these factors, the Fe(biphenolate)$_{3}$ network may have more advantages for producing the QAH effect than V$_2$O$_3$ in experiments.
In addition, a $2$D indium-phenylene organometallic framework (IPOF), In(C$_6$H$_5$)$_3$, also has a similar four-band character~\cite{PRL110.106804}. Differences from the Fe(biphenolate)$_{3}$ network are as follows: (i) these bands are due to In-p$_{x,y}$ orbitals, (ii)~many other bands cross them, and (iii) the FM ordering makes the band structures more complex. Hence, three QAH gaps predicted for the Fe(biphenolate)$_{3}$ network do not exist for IPOF.

Our calculation results show that the ground state of the Fe(biphenolate)$_{3}$ network is not the FM state, but a N\'{e}el antiferromagnetic (NAFM) state, which can be explained by a superexchange interaction mechanism based on a considerable hybridization of Fe-$3$d and biphenolate-$2$p orbitals. However, the NAFM ground state is a normal antiferromagnetic insulator, and there are no nontrivial features near the Fermi level (see the Supplementary Information for more details~\cite{SM}). Therefore, how to realize the FM ordering is an important issue. An approach that can be achieved experimentally is to use a high temperature and an external magnetic field, as follows: (i) Raise the temperature to transition the system into a \mbox{(near-)}paramagnetic state. (ii) Apply a magnetic field and decrease the temperature. This pins the system in the FM ordering.
A second approach is to use a FM substrate to make the Fe(biphenolate)$_{3}$ network in a FM ordering during sample synthesis. Obviously, the second method is more challenging, as the FM substrate must satisfy multiple demanding conditions, such as being an insulator and having a small mismatch, a small distance between surface atoms, and weak interactions with the Fe(biphenolate)$_{3}$ network.

In conclusion, we have systematically studied the $2$D MOM Fe(biphenolate)$_{3}$ network. Although its ground state is a NAFM insulator, when the material is transformed to a FM metastable state, it exhibits the QAH effect with three consecutive nontrivial gaps. This result was determined by calculating the Hall conductivity, edge states, and Berry curvatures. We confirmed that the nontrivial topology is from the two degenerate orbitals Fe-3d$_{xz}$ and 3d$_{yz}$. We also studied the role of electronic correlation in the topological and magnetic properties, finding that it does not cause qualitative changes. In addition, the effect of substrates was investigated in detail. Our results show that the experimentally used Ag($111$) and Cu($100$) substrates are not conducive to producing the QAH effect due to their good metallicity. \textit{h}-BN is a promising candidate. It ensures the Fe(biphenolate)$_{3}$ network stays flat and preserves the three QAH gaps. Topological edge states can be observed experimentally when the Fermi level is shifted into any one of the QAH gaps. Our results may provide useful information for experimental observations of the QAH effect in $2$D MOMs.

\section{Methods}
\label{part:methods}
\subsection{First-principles methods}
We used the density-functional-theory (DFT) code Vienna ab initio simulation package (\textsc{vasp}) to optimize the crystal structure and to calculate the basic electronic properties~\cite{PRB54.11169,CMS6.15,PRB47.558}. The self-consistent calculations were done using a $k$-mesh of size $5 \times 5 \times 1$ for the freestanding Fe(biphenolate)$_3$ network and $3 \times 3 \times 1$ for the system with a substrate. The energy cutoff was fixed at $500$~eV, and the tolerance for the total energy was smaller than $10^{-4}$~eV. Exchange interactions and the electronic correlation were treated within the generalized gradient approximation of Perdew, Burke, and Ernzerhof~\cite{PRL77.3865}.

\subsection{Four-band tight-binding model}\label{subsec:TBmodel}
In the FM state of the Fe(biphenolate)$_3$ network, there are four bands due to the degenerate 3d$_{xz}$ and 3d$_{yz}$ orbitals of Fe atoms near the Fermi level.
We employ a four-band tight-binding model to fit the four bands.
The NN hopping Hamiltonian is written as
\begin{equation}
{H_{NN}} = t\sum\limits_{i,j} {\left( {d_{j,{R_i}}^\dag {d_{j,{R_i} + {\delta _j}}} + h.c.} \right)},
\end{equation}
where $t$ is the NN hopping parameter, $\mathbf{R}_i$ and $\mathbf{R}_i+\mathbf{\delta}_j$, respectively, correspond to the A and B sublattices of Fe, $\mathbf{\delta}_j$ with $j = 1$, 2, 3, is the NN vector from the A site to the B site [Fig.~\ref{fig:struct}(b)], and ${d_{j,r}} = \left( {{d_{xz,r}}\hat x + {d_{yz,r}}\hat y} \right) \cdot {\hat \delta _j}$, where $\hat \ $ denotes a unit vector.
After the Fourier transform
\[
{d_R} = \frac{1}{{\sqrt N }}\sum\limits_k {{e^{ikR}}{d_k}},
\]
we have
\begin{equation}
  H_{NN} = t\sum\limits_k {{{\left( {\begin{array}{*{20}{c}}
{d_{xz,A,k}^\dag }\\
{d_{yz,A,k}^\dag }\\
{d_{xz,B,k}^\dag }\\
{d_{yz,B,k}^\dag }
\end{array}} \right)}^T}\left( {\begin{array}{*{20}{c}}
0&0&A_{k}&B_{k}\\
0&0&B_{k}&C_{k}\\
A^{*}_{k}&B^{*}_{k}&0&0\\
B^{*}_{k}&C^{*}_{k}&0&0
\end{array}} \right)\left( {\begin{array}{*{20}{c}}
{{d_{xz,A,k}}}\\
{{d_{yz,A,k}}}\\
{{d_{xz,B,k}}}\\
{{d_{yz,B,k}}}
\end{array}} \right)}
\end{equation}
with
\begin{align*}
A_{k} &= {\frac{3}{4}{e^{ik{\delta _1}}} + \frac{3}{4}{e^{ik{\delta _2}}}}, \\
B_{k} &= {\frac{{\sqrt 3 }}{4}{e^{ik{\delta _1}}} - \frac{{\sqrt 3 }}{4}{e^{ik{\delta _2}}}}, \\
C_{k} &= {\frac{1}{4}{e^{ik{\delta _1}}} + \frac{1}{4}{e^{ik{\delta _2}}} + {e^{ik{\delta _3}}}}.
\end{align*}

Similarly, we can obtain the NNN hopping Hamiltonian matrix in momentum space:
\begin{equation}
H_{NNN} = t'\sum\limits_k {{{\left( {\begin{array}{*{20}{c}}
{d_{xz,A,k}^\dag }\\
{d_{yz,A,k}^\dag }\\
{d_{xz,B,k}^\dag }\\
{d_{yz,B,k}^\dag }
\end{array}} \right)}^T}\left( {\begin{array}{*{20}{c}}
A'_{k}&B'_{k}&0&0\\
B'_{k}&C'_{k}&0&0\\
0&0&A'_{k}&B'_{k}\\
0&0&B'_{k}&C'_{k}
\end{array}} \right)\left( {\begin{array}{*{20}{c}}
{{d_{xz,A,k}}}\\
{{d_{yz,A,k}}}\\
{{d_{xz,B,k}}}\\
{{d_{yz,B,k}}}
\end{array}} \right)}
\end{equation}
with
\begin{align*}
A'_k &= {\frac{1}{2}\left[ \begin{array}{l}
\cos \left( {k{\kappa _1}} \right) + 4\cos \left( {k{\kappa _2}} \right) + \cos \left( {k{\kappa _3}} \right)
\end{array} \right]}, \\
B'_k &= {\frac{{\sqrt 3 }}{2}\left[ {\cos \left( {k{\kappa _1}} \right) - \cos \left( {k{\kappa _3}} \right)} \right]}, \\
C'_k &= {\frac{3}{2}\left[ {\cos \left( {k{\kappa _1}} \right) + \cos \left( {k{\kappa _3}} \right)} \right]},
\end{align*}
where $t'$ is the NNN hopping parameter and $\kappa_j$ with $j = 1$, 2, 3, are three of the six NNN vectors shown in Fig.~\ref{fig:struct}(b).

Spin--orbit coupling (SOC), as a relativistic correction to the Schr\"{o}dinger equation, plays an important role in driving topological phase transitions. In the current system, the SOC is described by
\begin{equation}
\begin{aligned}
{H_{SOC}} &= i\lambda \sum\limits_i
{\left( \begin{array}{l}   
d_{xz,{R_{A,i}}}^\dag {d_{yz,{R_{A,i}}}} - d_{yz,{R_{A,i}}}^\dag {d_{xz,{R_{A,i}}}}\\
 + d_{xz,{R_{B,i}}}^\dag {d_{yz,{R_{B,i}}}} - d_{yz,{R_{B,i}}}^\dag {d_{xz,{R_{B,i}}}}
\end{array} \right)} \\
   &= {i\lambda }\sum\limits_k {{{\left( {\begin{array}{*{20}{c}}
{d_{xz,A,k}^\dag }\\
{d_{yz,A,k}^\dag }\\
{d_{xz,B,k}^\dag }\\
{d_{yz,B,k}^\dag }
\end{array}} \right)}^T}\left( {\begin{array}{*{20}{c}}
0&1&0&0\\
{ - 1 }&0&0&0\\
0&0&0&1\\
0&0&{ - 1 }&0
\end{array}} \right)\left( {\begin{array}{*{20}{c}}
{{d_{xz,A,k}}}\\
{{d_{yz,A,k}}}\\
{{d_{xz,B,k}}}\\
{{d_{yz,B,k}}}
\end{array}} \right)}.
\end{aligned}
\end{equation}

In addition, we also introduce the chemical potential terms to realize the band fitting to the DFT results:
\begin{equation}
\begin{aligned}
{H_{CP}} &= \sum\limits_i {\left( \begin{array}{l}
{\varepsilon _{xz,A}}d_{xz,{R_{A,i}}}^\dag {d_{xz,{R_{A,i}}}} + {\varepsilon _{yz,A}}d_{yz,{R_{A,i}}}^\dag {d_{yz,{R_{A,i}}}}\\
 + {\varepsilon _{xz,B}}d_{xz,{R_{B,i}}}^\dag {d_{xz,{R_{B,i}}}} + {\varepsilon _{yz,B}}d_{yz,{R_{B,i}}}^\dag {d_{yz,{R_{B,i}}}}
\end{array} \right)} \\
  &= \sum\limits_k {{{\left( {\begin{array}{*{20}{c}}
{d_{xz,A,k}^\dag }\\
{d_{yz,A,k}^\dag }\\
{d_{xz,B,k}^\dag }\\
{d_{yz,B,k}^\dag }
\end{array}} \right)}^T}\left( {\begin{array}{*{20}{c}}
{{\varepsilon _{xz,A}}}&0&0&0\\
0&{{\varepsilon _{yz,A}}}&0&0\\
0&0&{{\varepsilon _{xz,B}}}&0\\
0&0&0&{{\varepsilon _{yz,B}}}
\end{array}} \right)\left( {\begin{array}{*{20}{c}}
{{d_{xz,A,k}}}\\
{{d_{yz,A,k}}}\\
{{d_{xz,B,k}}}\\
{{d_{yz,B,k}}}
\end{array}} \right)}.
\end{aligned}
\end{equation}
Since two sublattices and two orbitals are degenerate, we set ${\varepsilon _{xz,A}} = {\varepsilon _{yz,A}} = {\varepsilon _{xz,B}} = {\varepsilon _{yz,B}} = {\varepsilon _0}$.

Then, the total Hamiltonian is given as
\begin{equation}
H = {H_{NN}} + {H_{NNN}} + {H_{SOC} + {H_{CP}}}.
\end{equation}

\begin{acknowledgments}
X.-L. Yu is supported by Natural Science Foundation of Guangdong Province (Grant No. 2023A1515011852). J. Wu is supported by Natural Science Foundation of Guangdong Province (Grants No. 2017B030308003, No. 2019B121203002) and Science, Technology and Innovation Commission of Shenzhen Municipality (Grants No. KYTDPT20181011104202253, No. JCYJ20170412152620376).
\end{acknowledgments}

\newpage
\onecolumngrid

\renewcommand{\thesection}{S-\arabic{section}}
\setcounter{section}{0}  
\renewcommand{\theequation}{S\arabic{equation}}
\setcounter{equation}{0}  
\renewcommand{\thefigure}{S\arabic{figure}}
\setcounter{figure}{0}  

\indent

\begin{center}\large
\textbf{Supplementary Information for ``Three consecutive quantum anomalous Hall gaps in a metal-organic network"}
\end{center}
\section{Comparison of band structures}
\begin{figure}[tbp]
\centering
\setlength{\abovecaptionskip}{2pt}
\setlength{\belowcaptionskip}{4pt}
\includegraphics[angle=0, width=0.6 \columnwidth]{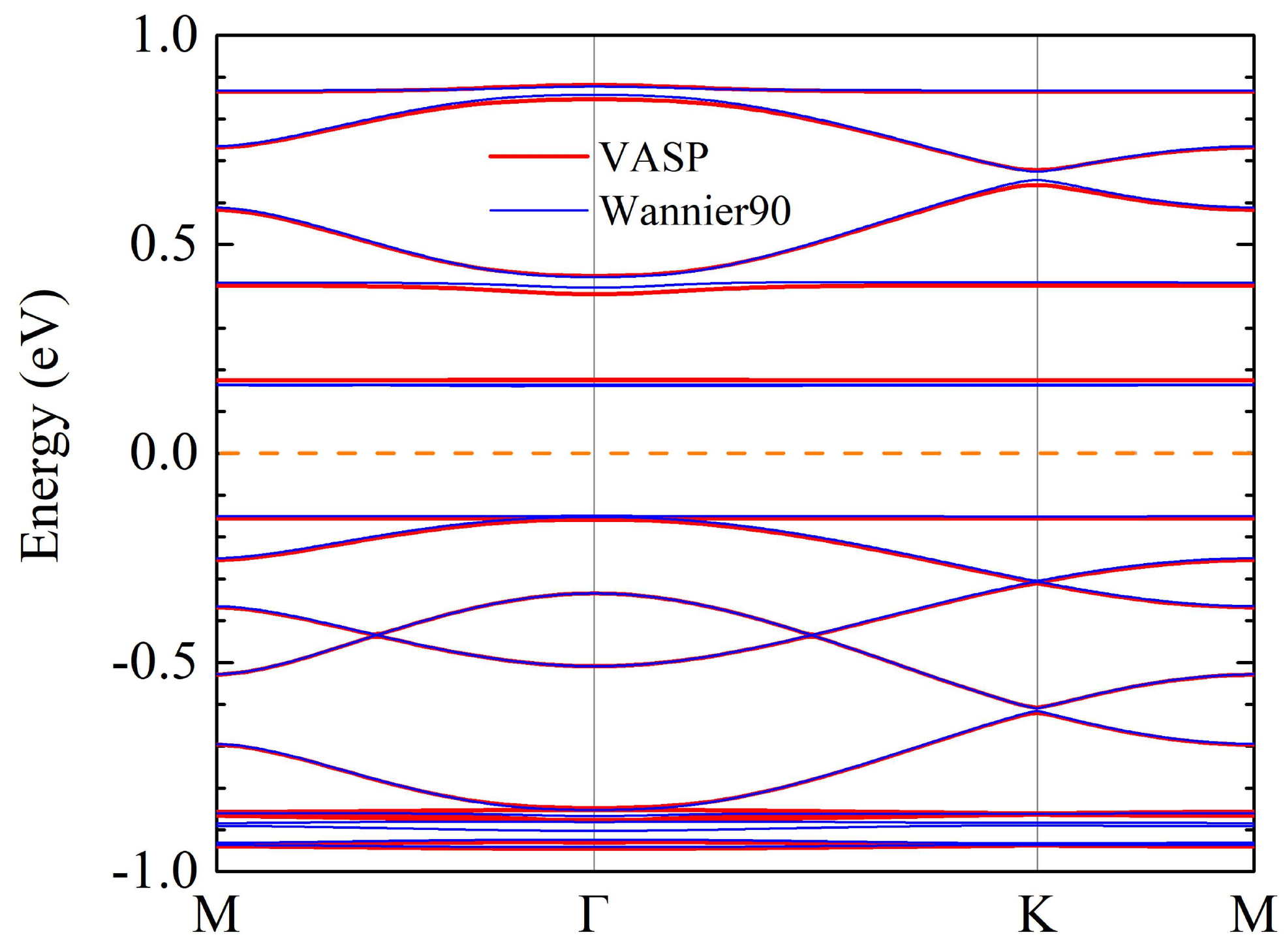}
\caption{Band structures of the FM state of the Fe(biphenolate)$_3$ network with SOC obtained from VASP (red) and Wannier$90$ (blue) codes.}
  \label{fig:VASP_Wannier90}
\end{figure}

We calculate the band structure of the ferromagnetic (FM) state of the Fe(biphenolate)$_3$ network with spin-orbit coupling (SOC) by using the density-functional-theory (DFT) code Vienna ab initio simulation package (\textsc{vasp})~\cite{PRB54.11169,CMS6.15,PRB47.558}. Then, according to the DFT result, we use the \textsc{Wannier}$90$ code to perform band fitting~\cite{CPC178.685}, where $272$ orbitals are considered, including Fe-3d, O-2p and C-2p orbitals. Their comparison is shown in Fig. ~\ref{fig:VASP_Wannier90} and the fitting near Fermi level is good.
We have used the fitting result to calculate Hall conductivity and edge states in the main text.

\section{Lattice parameters, ground state, and effective magnetic interactions}\label{sec:}
\begin{figure}
\centering
\setlength{\abovecaptionskip}{2pt}
\setlength{\belowcaptionskip}{4pt}
  \includegraphics[width=0.6 \columnwidth]{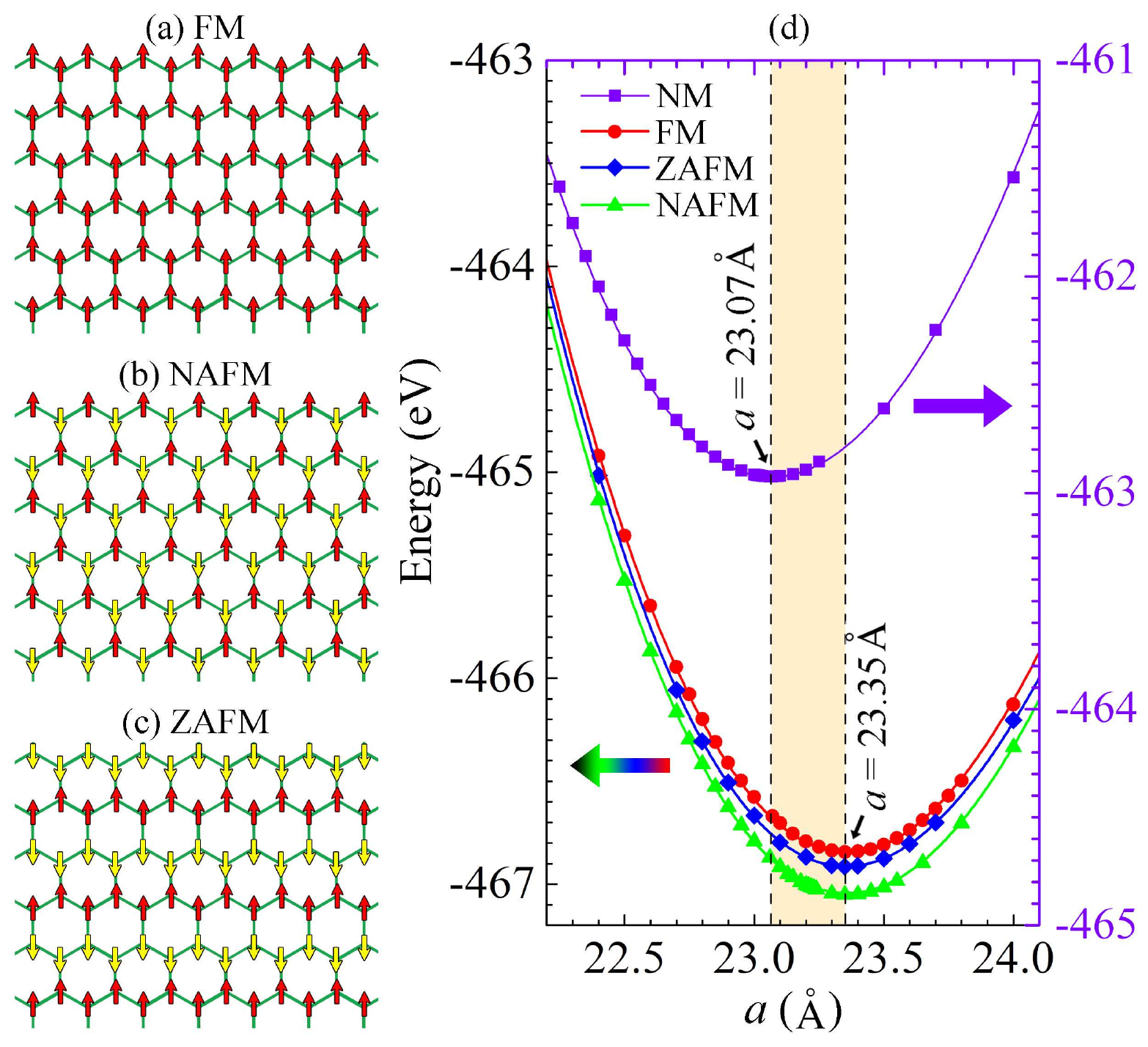}
  \caption{(a) FM, (b) NAFM, and (c) ZAFM configurations. (d) Energies of the four magnetic configurations as functions of the lattice parameter $a$.}
  \label{fig:3config}
\end{figure}

\begin{figure}[tbp]
\centering
\setlength{\abovecaptionskip}{2pt}
\setlength{\belowcaptionskip}{4pt}
\includegraphics[angle=0, width=0.95 \columnwidth]{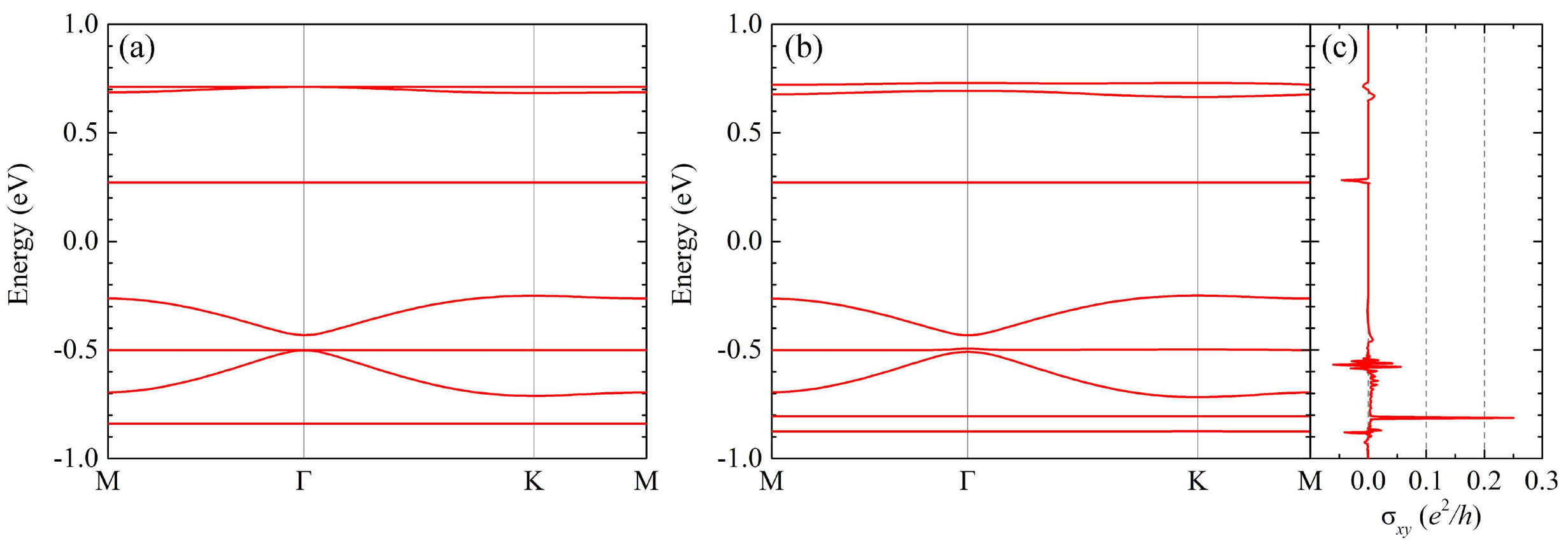}
\caption{Band structures of the NAFM state (a) without and (b) with SOC. (c) Hall conductivity of the NAFM state with SOC.}
  \label{fig:NAFM}
\end{figure}

To study the magnetic interactions between Fe atoms, we investigate three stable or metastable magnetic configurations: the FM, N\'{e}el antiferromagnetic (NAFM), and zigzag antiferromagnetic (ZAFM) orderings, as plotted in Figs. \ref{fig:3config}(a)--\ref{fig:3config}(c).
Figure \ref{fig:3config}(d) shows the total energies of the three magnetic states as functions of the lattice parameter $a$. All their lowest energies are at $a = 23.35$~\AA. Those of the NAFM and ZAFM states are lower than that of the FM state, and the NAFM state is the ground state.
The corresponding band structures without and with SOC are plotted in Figs. \ref{fig:NAFM}(a) and \ref{fig:NAFM}(b). We can see that the SOC opens several band gaps near the Fermi level. To check whether there exist topologically nontrivial states within these gaps, we further calculate Hall conductivity. However, we do not observe any quantization feature, as shown in Fig. \ref{fig:NAFM}(c), suggesting that the ground state is a normal NAFM insulator.
In addition, we plot the energy of the nonmagnetic (NM) state in Fig. \ref{fig:3config}(d) for comparison. It is much higher than the former three magnetic states. Its lowest-energy lattice parameter of $a = 23.07$~\AA{} is also different from that of the other configurations. We investigated the electronic and topological properties of the FM state in the parameter range marked by orange in Fig. \ref{fig:3config}(d) and found that there were no qualitative changes. Compared with the experimental data (nearest-neighbor (NN) Fe--Fe separation of approximately $a_{NN} = 13$~\AA{})~\cite{ACIE46.710}, the DFT optimized value $a_{NN} = 13.32$--$13.48$~\AA{} in this range is close to the experimental result. Note that during the optimization of the freestanding Fe(biphenolate)$_3$ network, the Fe-O-C angle remains $180^{\circ}$ without the clockwise/anticlockwise folding observed in the experiment. Moreover, all the atoms are in the same plane.

To gain an insight into the magnetic interactions, we employ the Heisenberg model to fit effective exchange integrals between magnetic atoms with an assumption that the energy differences between different magnetic configurations are predominantly due to interactions between magnetic atoms. Here, we consider NN and next-nearest-neighbor (NNN) exchange interactions. The corresponding Hamiltonian is
\begin{equation}\label{eq:Heisenberg}
H = {J_1}\sum\limits_{\left\langle {i,j} \right\rangle } {{\mathbf{S}_i} \cdot {\mathbf{S}_j}}  + {J_2}\sum\limits_{\left\langle {\left\langle {m,n} \right\rangle } \right\rangle } {{\mathbf{S}_m} \cdot {\mathbf{S}_n}}.
\end{equation}
$J_1$ and $J_2$ are NN and NNN exchange integrals, respectively. Correspondingly, ${\left\langle {i,j} \right\rangle }$ and ${\left\langle {\left\langle {m,n} \right\rangle } \right\rangle }$ denote summation over the NN and NNN sites, respectively.

Since all these are insulators, we can use the Heisenberg model to fit the exchange integrals. According to Eq.~(\ref{eq:Heisenberg}), their energy formulas are
\begin{equation}
\left\{
\begin{aligned}
{E_{FM}} &= 3{J_1}{S^2} + 6{J_2}{S^2} + {E_{other}}\\
{E_{NAFM}} &=  - 3{J_1}{S^2} + 6{J_2}{S^2} + {E_{other}}\\
{E_{ZAFM}} &= {J_1}{S^2} - 2{J_2}{S^2} + {E_{other}}
\end{aligned} \right.
\end{equation}
where $E_{FM}$, $E_{NAFM}$, and $E_{ZAFM}$ are, respectively, the total energies of the three magnetic configurations, and $E_{other}$ is the sum of all energies except the magnetic interactions. Using the total energy and magnetic moment obtained from the first-principles calculations, we can qualitatively estimate the effective exchange integrals $J_1$ and $J_2$.

Without considering the electronic correlation $U$ or SOC, the magnetic moment of an Fe atom is about $3.6 \mu_B$. Accordingly, if the spin is $S = 1.8$, then the exchange integrals ${J_1} = 6.235$~meV and ${J_2} = -1.123$~meV.
When SOC is considered, we obtain ${J_1} = 6.226$~meV and ${J_2} = -1.119$~meV.
The exchange integrals in the second case are smaller than those in the first case due to the SOC, but the difference between them is very small and can be neglected.

\begin{figure}
\centering
\setlength{\abovecaptionskip}{2pt}
\setlength{\belowcaptionskip}{4pt}
  \includegraphics[width=0.5 \columnwidth]{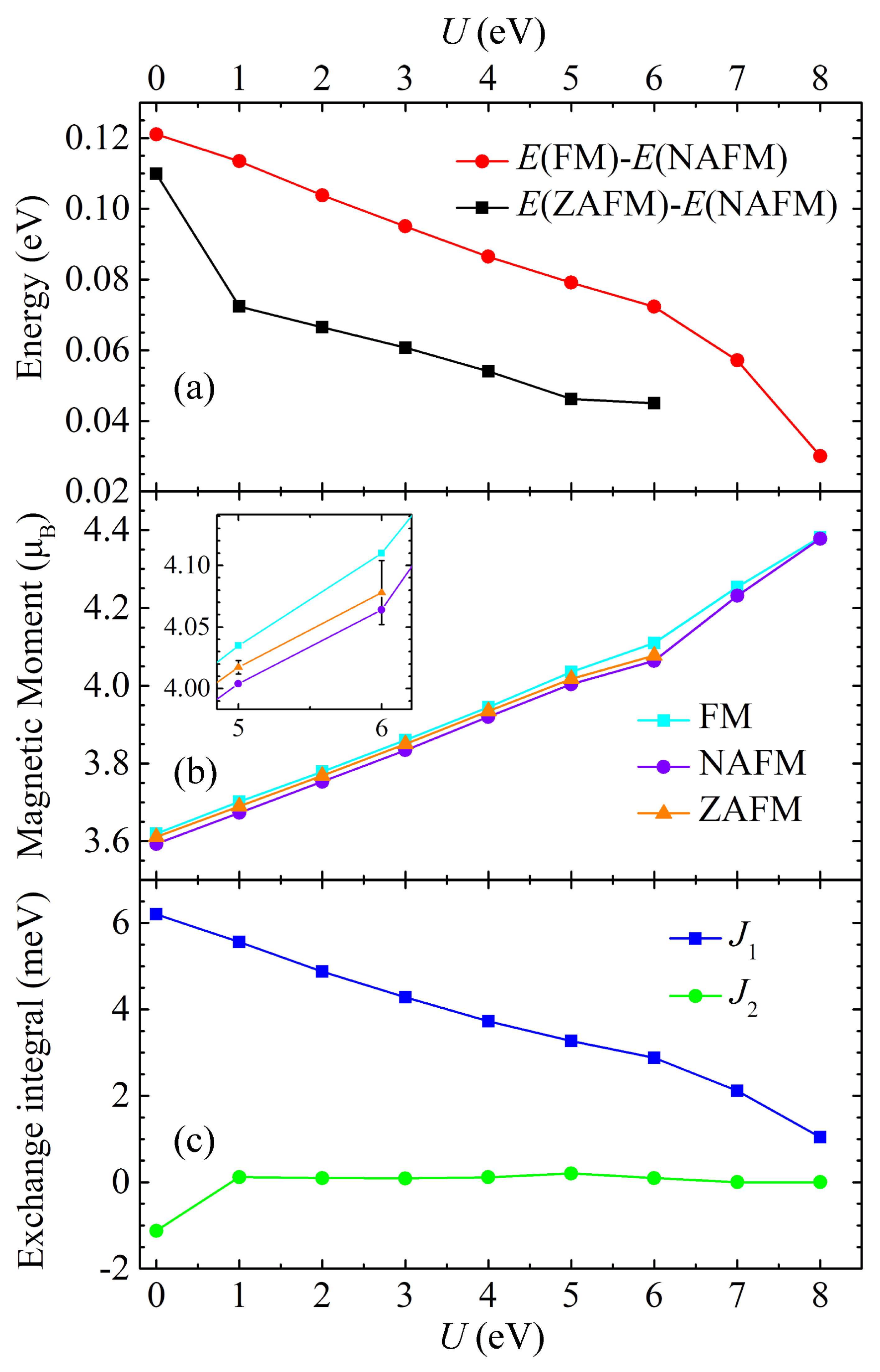}
  \caption{(a) Energy differences, (b) magnetic moments, and (c) exchange integrals as functions of $U$. Inset of panel (b) is an enlargement for the range $U = 5$--$6$~eV. The error bars indicate the instability of the magnetic moment in the ZAFM state.}
  \label{fig:E_M_J-U}
\end{figure}

We next calculate cases with different electronic correlations.
Figure \ref{fig:E_M_J-U} shows the energy differences, magnetic moments, and exchange integrals as functions of $U$. The above case without electronic correlation is also included for comparison. With increasing $U$, the energies of the FM and ZAFM states gradually become closer to that of the NAFM ground state. When $U > 6$~eV, the ZAFM state is unstable.
Figure \ref{fig:E_M_J-U}(b) shows that the spin magnetic moment of Fe for the three magnetic states gradually increases with an increase of $U$, which is reasonable because electron repulsion tends to align the electrons of each Fe atom in the same direction. In addition, the strength of the exchange integrals $J_1$ and $J_2$ decreases. The physical reason is that the exchange probability between neighboring Fe atoms decreases due to Coulomb repulsion.
In particular, for $U > 0$, the strength of $J_2$ dramatically decreases to a value close to zero.

\begin{figure*}
\centering
\setlength{\abovecaptionskip}{2pt}
\setlength{\belowcaptionskip}{4pt}
  \includegraphics[width=0.9 \columnwidth]{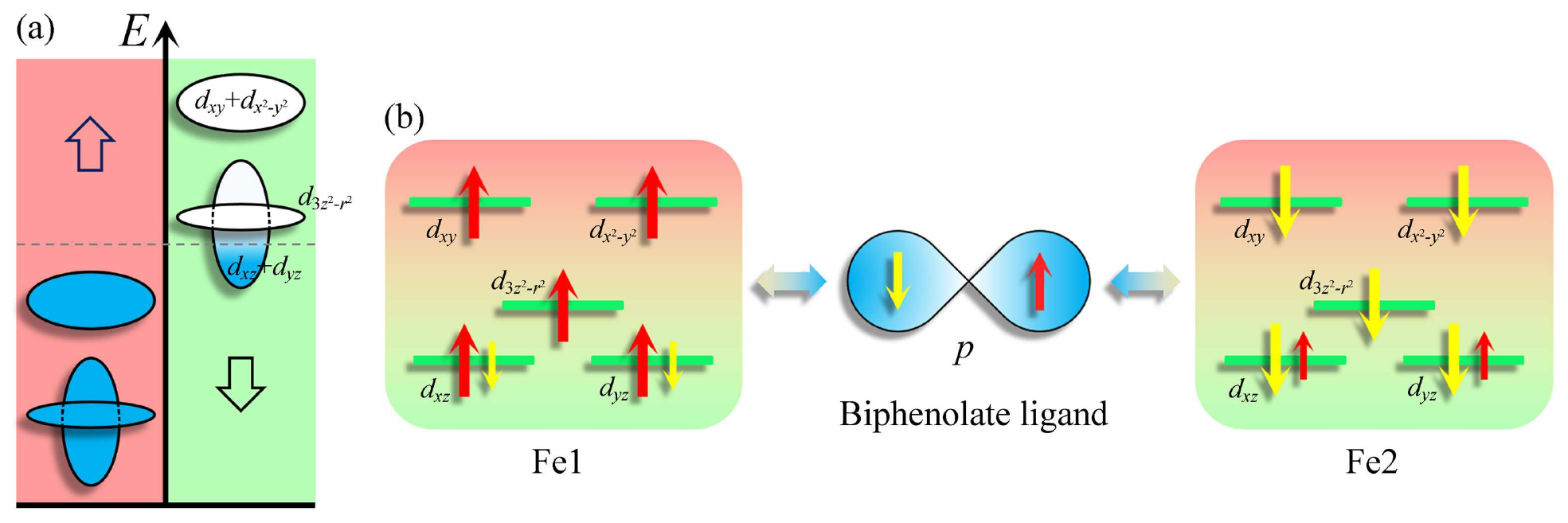}
  \caption{(a) Electronic configuration of Fe-3d orbitals. Red and green regions indicate spin up and down, respectively. (b) Schematic of the superexchange interaction for the NAFM ground state.}
  \label{fig:superexchange}
\end{figure*}

The physical mechanism of the NAFM ground state can be explained by a superexchange interaction based on a considerable hybridization of Fe-$3$d and biphenolate-$2$p orbitals.
In the ground state, when the spin-up (down) states of Fe-$3$d orbitals are fully occupied, the spin-down (up) states are partially occupied by 3d$_{xz}$ and 3d$_{yz}$, as illustrated in Fig. \ref{fig:superexchange}(a), resulting in a magnetic moment of each Fe of about $3.5$--$4\mu_B$. Figure \ref{fig:superexchange}(b) is a schematic diagram of the physical mechanism of the ground state.
When the magnetic moment of Fe$1$ is up, the up states of the five $3$d orbits are fully occupied, and the down states are partially occupied. Moreover, the spin-down electrons in the p orbitals of the biphenolate can hop to the $3$d orbitals of Fe$1$.
Similarly, for Fe$2$ with spin-down polarization, the spin-up electrons in the p orbitals of the biphenolate can hop to the $3$d orbitals of Fe$2$. Finally, the NAFM ground state forms due to the superexchange interaction between neighboring Fe atoms.

\section{Results for the chiral structure}
\begin{figure}[tbp]
\centering
\setlength{\abovecaptionskip}{2pt}
\setlength{\belowcaptionskip}{4pt}
\includegraphics[angle=0, width=0.9 \columnwidth]{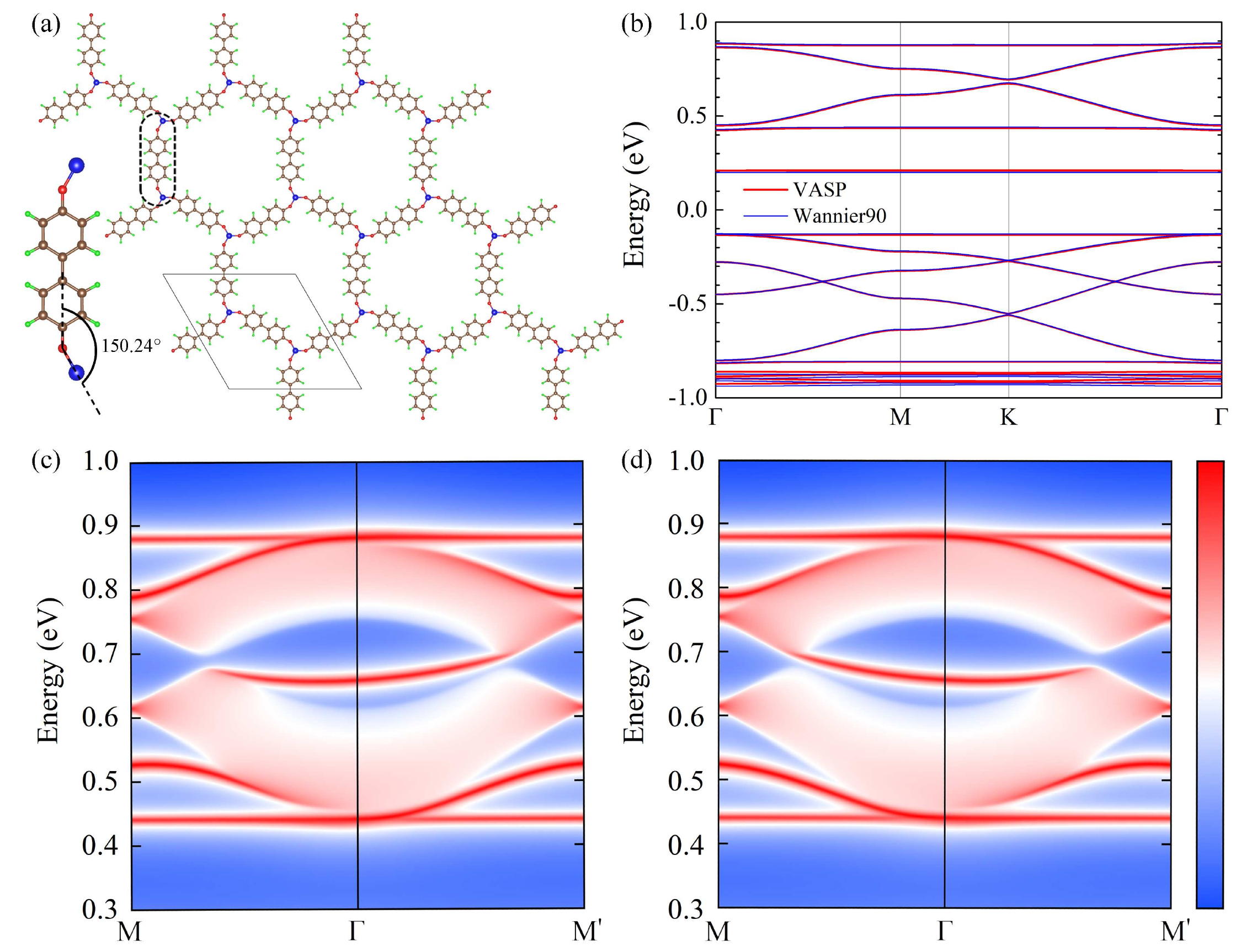}
\caption{(a) Lattice structure of the Fe(biphenolate)$_3$ network with a Fe-O-C angle of $150.24^{\circ}$. Blue, red, brown, and green spheres denote Fe, O, C, and H atoms. (b) Band structures of the FM state with SOC obtained from VASP (red) and Wannier$90$ (blue) codes. Corresponding band structures of semi-infinite systems with (c) left and (d) right edges. A high (low) density of states is coloured red (blue).}
  \label{fig:chirality}
\end{figure}

We consider the two-dimensional ($2$D) chirality of experimental samples with a threefold rotation symmetry~\cite{ACIE46.710}, which may be induced by substrates. However, the specific lattice structure was not given in experiments. Here, we artificially modify the free-standing lattice structure by rotating the three O atoms around the Fe atom with a Fe-O-C angle of $150.24^{\circ}$ [Fig. \ref{fig:chirality}(a)], and maintaining all bond lengths unchanged. The lattice parameter is $a=22.273 {\AA}$, and the nearest-neighbor (NN) Fe-Fe distance is $a_{NN} = 12.86 {\AA}$ which is close to the experimental value ($\sim 13 \AA$).
To check whether the nontrivial topology still exists in the chiral structure, we repeat the calculation process of the FM state. The DFT results show that the magnetic moment of each Fe is $3.62$ $\mu_B$, the energy gap is about $0.34$ eV, and there are several flat bands near the Fermi level, as shown in Fig. \ref{fig:chirality}(b). In the range of $0.4 \sim 0.9$ eV, the SOC significantly open three band gaps. Furthermore, based on the band fitting by Wannier$90$ code~\cite{CPC178.685}, semi-infinite systems are calculated using the \textsc{WannierTools} software~\cite{CPC224.4059}.
Figures \ref{fig:chirality}(c) and \ref{fig:chirality}(d) show that there are edge states within the three gaps and left- and right-edge states propagate along opposite directions, corresponding the quantum anomalous Hall (QAH) effect.
It can be seen that these results are very similar to those obtained in the optimized structure. Therefore, our current calculation results indicate that the small in-plane lattice distortion does not change the topologically nontrivial feature.

\section{Effect of Ag($111$) substrate}
\begin{figure}[tbp]
\centering
\setlength{\abovecaptionskip}{2pt}
\setlength{\belowcaptionskip}{4pt}
\includegraphics[angle=0, width=0.95 \columnwidth]{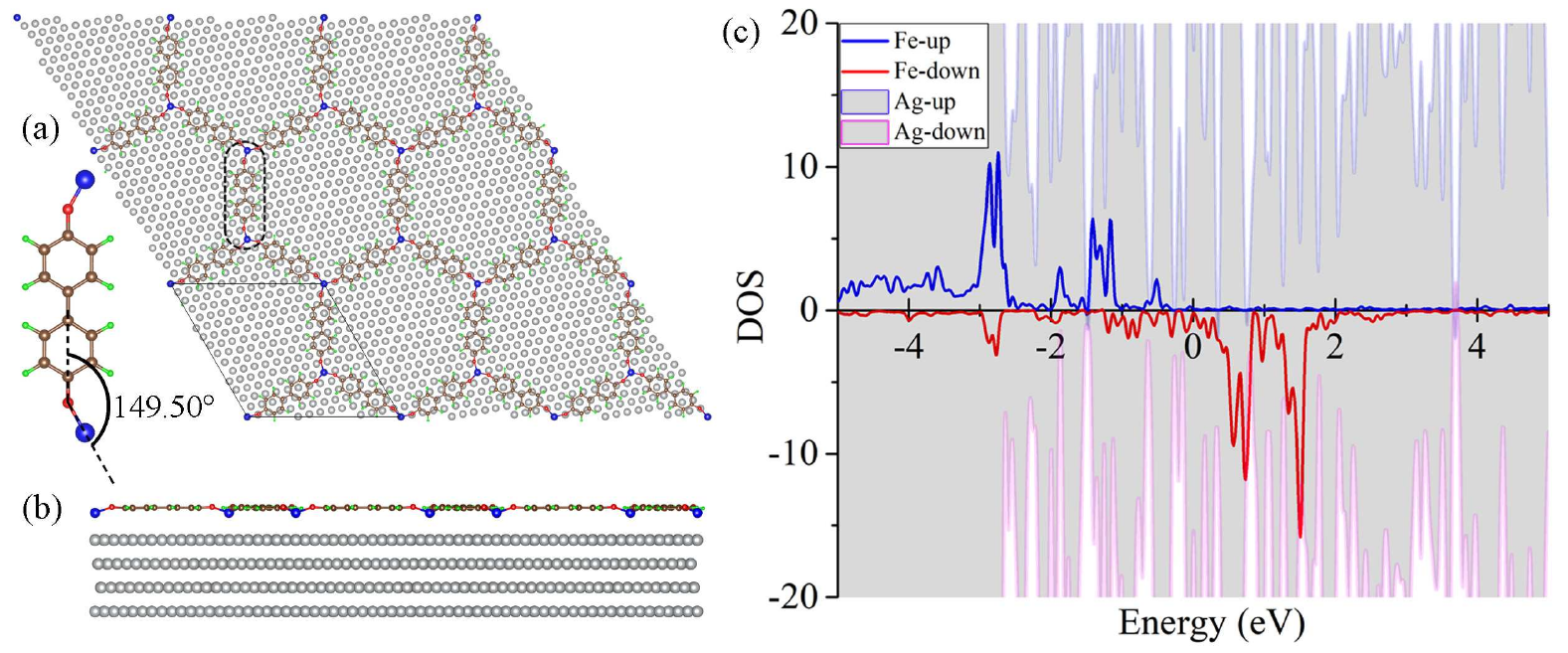}
\caption{(a) Top and (b) side views of the atomic configuration of the Fe(biphenolate)$_{3}$/Ag($111$) hybrid system. The Fe-O-C angle is $149.50^{\circ}$. (c) PDOS of the hybrid system in the FM states.}
  \label{fig:Ag}
\end{figure}

Figures \ref{fig:Ag}(a) and \ref{fig:Ag}(b) exhibit the optimized structure of the Fe(biphenolate)$_3$/Ag($111$) hybrid system, which has been synthesized in experiments~\cite{ACIE46.710}.
The optimization result shows an obvious $2$D chirality with a clockwise or anticlockwise structure around Fe atoms. It is in agreement with experimental measurements~\cite{ACIE46.710}. Besides, we also find that Fe atoms obviously deviate from the biphenolate plane.
It is attributed to the large atomic distance on the Ag($111$) surface.
Figure \ref{fig:Ag}(c) shows the projected density of states (PDOS) of the optimized hybrid system. One can see that the DOS of the Fe(biphenolate)$_3$ network is completely covered by that of the Ag($111$) substrate near the Fermi level. It can be attributed to the good metallicity of Ag.

\end{document}